\documentclass[aps, prl, groupedaddress, twocolumn]{revtex4-1}

\usepackage{amsmath}
\usepackage{graphicx}
\usepackage{subcaption}
\usepackage{color}

\usepackage[abs]{overpic}

\renewcommand{\[}{
\begin{equation}
\begin{aligned}
}
\renewcommand{\]}{
\end{aligned}
\end{equation}
}
\renewcommand{\endtitlepage}{\let\endtitlepage\relax}

\renewcommand{\vec}{\pmb}

\newcommand{\density}{\rho}

\newcommand{\refeq}[1]{Eq.~\ref{#1}}
\newcommand{\refpic}[1]{Fig.~\ref{#1}}

\newcommand{\vecr}{\vec{r}}
\newcommand{\vecrone}{\vec{r}_1}
\newcommand{\vecrtwo}{\vec{r}_2}

\newcommand{\enot}{{E_0}}
\newcommand{\enotstar}{\enot^*}

\newcommand{\vece}{\vec{E}}

\newcommand{\vectheta}{\vec{\theta}}

\newcommand{\norm}[1]{\left\|#1\right\|}

\newcommand{\densityr}{\density_{r}}
\newcommand{\densitytheta}{\density_{\theta}}
\newcommand{\densityg}{\density_{\gamma}}

\newcommand{\constc}{\mathcal{C}}

\newcommand{\conj}[1]{#1^*}

\newcommand{\twopictures}[5]
{
\begin{figure}
\centering
\begin{subfigure}{.25\textwidth}
  \centering
  \includegraphics[width=#3cm]{#1}
\end{subfigure}%
\begin{subfigure}{.25\textwidth}
  \centering
  \includegraphics[width=#3cm]{#2}
\end{subfigure}
\caption{#4}
\label{two_pictures_#5}
\end{figure}
}

\newcommand{\sigmasx}{\sigma_{sx}}
\newcommand{\sigmagx}{\sigma_{gx}}

\newcommand{\csd}{W}
\newcommand{\csdop}{A_\csd}
\newcommand{\sdc}{\mu}
\newcommand{\freq}{\omega}
\newcommand{\eigval}{\lambda}
\newcommand{\cohmode}{\phi}
\newcommand{\ngrid}{N_{grid}}
\newcommand{\nmode}{N_m}
\newcommand{\nx}{N_x}
\newcommand{\ny}{N_y}
\newcommand{\cohf}{CF}

\begin{document}

\title{Coherent modes of X-ray beams emitted by undulators in new storage rings}

\author{Mark Glass}
\affiliation{ESRF - The European Synchrotron \\
71, Avenue des Martyrs, Grenoble, France}
\author{Manuel Sanchez del Rio}
\affiliation{ESRF - The European Synchrotron \\
71, Avenue des Martyrs, Grenoble, France}

\date{\today}

\begin{abstract}
Synchrotron radiation emitted by electrons passing through an undulator placed in a storage ring 
is decomposed in coherent modes. The case of ultimate storage rings where the electron emittance
is comparable to the emittance of the photon fan is analyzed by means of the cross spectral density
and the coherent mode spectrum. 
The proposed method permits naturally the statistical analysis and propagation of the cross spectral 
density along the beamline optics. The coherence properties of the X-ray beam at
any point of the beamline are completely given in terms of the eigenvalues and
coherent modes of the cross spectral density.
\end{abstract}

%\keywords{}

\maketitle

% \section{Introduction **section titles to be removed**}

Synchrotron radiation (SR) has witnessed an enormous growth in the last decades
because of its applicability to multidisciplinary applied science. Several generations of synchrotron radiation
sources (first generation or application of SR in parasitic mode, second generation or storage
rings dedicated to SR, third generation or storage rings with long straight sections to hold
undulators) have raised the brilliance with rates even higher than the Moore law. The current third
generation of synchrotrons is now continued in two different new facilities: X-ray free electron lasers (XFELs), 
based on linear accelerator technology, and the so-called ``diffraction 
limited storage rings''. These are circular storage rings
where the electron horizontal emittance is lowered to a level comparable to the
present vertical emittance, which is small enough to allow experiments
exploiting the X-ray beam coherence.
In this context, many experimental techniques have developed,
like X-ray photon correlation spectroscopy, coherent diffraction imaging, or ptychography. In all
new facilities or upgrades of the existing ones, the keyword ``coherence'' is omnipresent. We 
look to the upgrade of the existing facilities, like the EBS (Extremely
Brilliant Source) at the European Synchrotron Radiation Facility, aiming at building a storage ring 
of 150 pm emittance (as compared with the present one of 4 nm) that will boost the X-ray brilliance and
the coherence properties.
The accurate calculation and quantitative evaluation of the parameters related to X-ray coherence in 
new storage rings is the subject of this paper. 

The calculation of observable quantities of the emission like the mean intensity
requires a statistical approach.
Kim \cite{kim1986} proposed a method to calculate the Wigner functions  
of synchrotron undulators for many practical cases. 
The method was revisited by Geloni {\textit{et al.}} in a
rigorous statistical optics context \cite{geloni2008}. One may define a 
virtual source in a plane before the beamline with the property that an electric field propagated from
the virtual source to a position along the beamline reproduces the real electric
field at that position. For sufficiently long electron beam bunches the storage
ring emission can be considered wide-sense stationary for many
practical applications \cite{geloni2008}. We can therefore describe second
order coherence phenomena in frequency domain with the cross spectral density
(CSD) $W(\vecrone, \vecrtwo, \freq)$ where $\vecrone$, $\vecrtwo$ are spatial
coordinates and $\freq$ is the radiation frequency.
If additionally the variation of the undulator magnetic field is negligible 
along the electron beam dimension, the cross spectral density at the virtual source is given by:
\[
\label{csd_synchrotron_general}
\csd(\vecrone, \vecrtwo, \freq)
=
N_e
\int
d\gamma
d\vecr
d\vec{\theta}
~
\density(\vecr, \vec{\theta},\gamma)
e^{ik\vec{\theta}(\vecrtwo-\vecrone)}                             \times     \\
\enotstar(\gamma,\vecrone-\vecr, \freq)
\enot(\gamma,\vecrtwo-\vecr, \freq)
\]
where $N_e$ is the number of electrons, $\density$ is the five-dimensional spatial,
angular and energy dependent electron phase space distribution, $\enot$ is the
reference electric field created by a reference electron, the star denotes
complex conjugation and $\gamma$ is the relativistic Lorentz factor of the
electrons.
The exact reference electric field $\enot$ produced by an undulator is given
by \cite{jackson}:
\[
\label{undulator_emission}
& \vece(\vec{R},\freq)
=
\frac{ie\freq}{4\pi c\epsilon_0}  \times \\
& \int 
\left[
\frac{\vec{n}\times[(\vec{n}-\dot\vecr)\times\ddot{\vecr}]}{(1-\dot\vecr
\vec{n})^2}
+ 
\frac{c}{\gamma^2R}
\frac{(n-\dot\vecr)}{(1-\dot\vecr\vec{n})^2}
\right]
e^{i\freq(t-\vec{n}\vec{r(t)}/c)}
dt 
\]
where $e$ is the electron charge, $c$ the velocity of light, $\epsilon_0$ the electric constant,
$\vec n(t)=\vec{R}-\vecr(t)/\norm{\vec{R}-\vecr(t)}$ is the unit vector
pointing from the particle to the observation point $\vec{R}$; $\vecr(t)$ is the
electron trajectory of the reference electron and the dot denotes the time
derivative.

The cross spectral density obeys the Wolf equation:
\[
\left[
\nabla^2_j
+k^2n^2(\vecr_j)
\right]
\csd(\vecrone, \vecrtwo,\freq)
=0,
\]
where $j=1,2$; $k$ is the wave number in free-space, and $n(\vecr)$ is the
refractive index. The Wolf equation resembles the Helmholtz
equation. Indeed, it can be seen as a tensor product version of the Helmholtz equation.
%===============================
With the Helmholtz equation, the usual
Green's function method for the propagation of wave fields can be
derived \cite{jackson}:
\[
\label{wave_propagation_green}
E'(\vecr, \freq)
=
\frac{1}{2\pi}
\int_{\mathcal A} d\vecr'
E(\vecr',\freq)
\frac{\partial}{\partial n'}
G(\vecr,\vecr'),
\]
where $\mathcal A$ is a closed surface on which $E$ is known, $\vec{n}$ is a
unit vector normal to $\mathcal A$, and $G$ is a Dirichlet Green's function for the wave equation
that satisfies appropriate boundary conditions. 
%=======================================
In consequence, the propagation formula for the cross spectral density takes the
form of a tensor product version of the propagation formula of wave fields:
\[
\label{csd_propagation_green}
\csd'(\vecrone, \vecrtwo, \freq)
=
\frac{1}{(2\pi)^2}
\int_{\mathcal A} d\vecrone'
\int_{\mathcal A} d\vecrtwo'  \times \\
W(\vecrone',\vecrtwo', \freq)
\frac{\partial}{\partial n'_1}
\conj{G}(\vecrone,\vecrone')
\frac{\partial}{\partial n'_2}
G(\vecrtwo,\vecrtwo'),
\]
where $\vec{n}_1$ and $\vec{n}_2$
are unit vectors normal to $\mathcal A$. 
The spectral density $S(\vecr, \freq)=\csd(\vecr, \vecr, \freq)$ can be
seen as the energy per unit time \cite{saleh2007fundamentals}. The spectral
degree of coherence is defined by \cite{mandel_wolf}:
\[
\sdc(\vecrone, \vecrtwo, \freq) = \frac{\csd(\vecrone, \vecrtwo,
\freq)}{\sqrt{S(\vecrone, \freq)S(\vecrtwo, \freq)}}.
\]
The cross spectral density can be represented in terms
of eigenvalues $\eigval_m(\freq)$ and coherent modes $\cohmode_m(\vecr, \freq)$
\cite{mandel_wolf}:
\[
W(\vecrone, \vecrtwo, \freq)
=
\sum_m
\eigval_m(\freq)
\conj\cohmode_m(\vecrone,\freq)
\cohmode_m(\vecrtwo, \freq).
\]
The eigenvalues and the coherent modes are the solution of the
homogeneous Fredholm integral equation of second kind:
\[
\label{fredholm_equation}
\int d\vecrone
W(\vecrone,\vecrtwo, \freq){\cohmode}_m(\vecrone,
\freq)=\eigval_m(\freq)\cohmode_m(\vecrtwo, \freq).
\]
This is an eigenvalue problem and may be rewritten as:
\[
\csdop[\cohmode_m]=\eigval_m \cohmode_m
\]
with the Hermitian integral operator $\csdop$:
\[
\csdop[f](\vecr)
=
\int d\vecr'~
W(\vecr',\vecr, \freq)f(\vecr').
\]
Without loss of generality, the coherent modes can be assumed to be orthonormal.
The sum of the eigenvalues equals the integrated spectral density.
We will use the notion used in the mathematical field of spectral theory and call the
set of the eigenvalues the eigenvalue spectrum. We define the mode distribution
as:
%$d_m(\freq)=\eigval_m(\freq)/\sum_n \eigval_n(\freq)$. 
\[
\label{spectrum}
d_m=\frac{\eigval_m}{\sum_n \eigval_n}.
\]
The index $m=0$ refers to the first coherent mode.

The coherent modes are statistically uncorrelated \cite{mandel_wolf}. The eigenvalue of a coherent mode can
therefore be seen as the spectral density carried by this mode and the
propagation of the cross spectral density can equally be expressed in terms of
propagated coherent modes. Let $\cohmode'_m$ denote the propagated coherent
mode as if it was an electric field.
The propagated cross spectral density $W'$ may be written as:
\[
\label{propagated_csd}
W'(\vecrone, \vecrtwo, \freq)
=
\sum_m
\eigval_m(\freq)
\conj{\left(\cohmode'_m(\vecrone,\freq)\right)}
\cohmode'_m(\vecrtwo,\freq).
\]

If the eigenvalue spectrum is narrow the propagation of the cross spectral
density by means of the propagation of the coherent modes is computationally
advantageous. Assume there are $\nmode$ coherent modes and the
sum of their eigenvalues approximate to a given level of accuracy (e.g. 99\%) the total spectral density.
Imagine the surface $\mathcal A$ in equation \refeq{csd_propagation_green} is spanned numerically by
a grid of $\ngrid$ points. If $\nmode \ll \ngrid$ then it is more favorable to
calculate \refeq{wave_propagation_green} for $\nmode$ modes than to calculate
\refeq{csd_propagation_green}.

A prominent model in statistical optics with known analytical coherent mode
decomposition is the Gaussian Schell-model (GSM).
It is used to model laser modes at the waist \cite{siegman1971}. It has been applied
to approximate undulator radiation in high emittance storage rings \cite{coisson1997,vartanyants2010,lindberg2015} 
The GSM supposes that the cross spectral density takes a Gaussian form defined by 
$\sigmasx(\freq)$ and $\sigmagx(\freq)$, the standard deviations of the spectral density and the
spectral degree of coherence, respectively. The interest of the model, in addition to its applicability
to describe laser radiation, resides in the fact that its decomposition in coherent modes can be done
analytically \cite{starikov1982, gori1980} leading to the Gaussian-Hermite eigenfunctions, with eigenvalues
that decrease exponentially, as described in the supplementary
material.
The electric field emitted by the single electron in an undulator is rather complex and in general a Gaussian
approximation is not justified. Significant deviations of the cross spectral density from 
a GSM will become apparent when the electron beam emittance becomes
comparable or smaller than the extent of the single electron radiation at the virtual source because radiation details are no longer smeared out by the electron beam. 
This situation is expected in the case of the new ultra low emittance storage rings.

% \section{Our method}

The electric field of the reference electron $\enot$ can be calculated
only approximately with rather complicated analytical formulas \cite{geloni2008,elleaume}
(usually only valid for photon
energies close to a harmonic) that lead to integral expressions for which no
closed solutions are known.
In consequence \refeq{csd_synchrotron_general} can not be solved in a closed form and numerical techniques must be applied.

The innovative aspect of this letter is the successful numerical realisation of
a coherent mode decomposition for undulator storage ring emission, i.e.
a numerical solution of \refeq{fredholm_equation}. The
reference electric field $\enot$  is calculated
by numerical integration of \refeq{undulator_emission} using SRW \cite{Chubar1998}. 
We solve the Fredholm equation \refeq{fredholm_equation}
numerically with no further approximation using a basis set consisting of
normalized two-dimensional step-functions. A
two-dimensional step-function is a function that is constant on a rectangle
and zero elsewhere. Hereby the center of the rectangle lies on a grid
point. 
The grids we use have constant step width in each direction so our step-functions come from the 
same mother step-function shifted to cover all the domain without overlapping.
The representing matrix of $\csdop$ in that
basis scales with $\nx^2\ny^2=\ngrid^2$ where $\nx$, $\ny$ are the numbers of
horizontal and vertical grid points, respectively.
This scaling behavior of the representing matrix leads quickly to
practically problematic memory requirements. We solve the eigenvalue equation with the
iterative eigensolver SLEPc~\cite{Hernandez2005}. The library SLEPc allows the
implementation of user-defined matrix-vector multiplications that are then used
by the eigensolver. We assume an electron phase space
density that separates into a spatial part $\densityr$, a divergence part
$\densitytheta$ and an energy part $\densityg$:
\[
% \begin{split}
\csdop(f)(\vecrtwo)=
\int
d\gamma
\densityg(\gamma)
\int
d\vecrone
d\vec{\theta}
~
\densitytheta(\vec{\theta})
e^{ik\vec{\theta}(\vecrtwo-\vecrone)} \times  \\
\int
d\vecr
\densityr(\vecr,\gamma)
\enotstar(\gamma,\vecrone-\vecr,\freq)
\enot(\gamma,\vecrtwo-\vecr,\freq)
f(\vecrone).
% \end{split}
\]
The last integral can be considered as an undulator CSD from
\refeq{csd_synchrotron_general} with a delta shaped divergence and a fixed
electron energy.
It can therefore be decomposed into eigenvalues $\alpha_n$ and coherent modes $\Psi$.
The $\vec{\theta}$ integration can be performed because it is a Fourier transform. The operator
$\csdop$ can now be rewritten as:
\[
\csdop(f)(\vecrtwo)=
\int
d\gamma
\densityg(\gamma)
\int
d\vecrone
\hat\densitytheta(k(\vecrtwo-\vecrone)) \times \\
\sum_n
\alpha_n
\conj\Psi_n(\vecrone)
\Psi_n(\vecrtwo)
f(\vecrone).
\]
For a few hundreds modes $\nmode$ this two-step approach can be performed
efficiently numerically. This approach can be easily generalized for electron
phase space densities that have an additional spatial and divergence
coupling term of the form $e^{\constc\vecr\vectheta}$ where $\constc$ is a
constant.
These coupling terms appear if the Twiss alpha value is non zero, i.e. at
positions of the lattice that are not a symmetry point. In many cases the memory consumption of the two-step approach is of the order of $\nmode \cdot \ngrid$ while the full representing matrix in the chosen basis set needs $\ngrid^2$ elements.
The GSM approximation motivates the expectation of a small number of
coherent modes for very low emittance storage rings. The number of
significant coherent modes depends strongly on the emittance of the lattice
and the energy of the undulator radiation. Typical values for $\nmode$
are a few hundreds to a few thousands and typical values for $\ngrid$ are a few hundred thousand.

The coherent modes can be propagated in free space using standard wave optics 
Fresnel propagation methods. The Gauss-Hermite modes in the Gaussian Schell-model, 
are invariant under propagation in the far field (Fraunhofer). The coherent
modes obtained for undulator radiation do not present, in general, this property.
We use SRW for the propagation of the modes. The propagated cross spectral
density $W'$ is easily deduced from the propagated modes using \refeq{propagated_csd}. 
The coherent modes change by propagation in free space, but the
eigenvalue spectrum remains constant.
An interesting case is the effect of a slit, pinhole, or a more general optical element. When the 
element cuts the beam (removes intensity), the eigenfunctions are clipped by a
window, so their norm becomes smaller than one. Therefore, the coherent modes
and the spectrum of coherent modes of $W'$ are different from those of $W$ and we
can solve \refeq{fredholm_equation} again to determine the new coherent modes
and eigenvalue spectrum of $W'$.

% \section{Applications}

Many tests were performed to verify our algorithm. They include delta
shaped electron beams, a comparison of the numerical decomposition with a
Gaussian reference electric field\label{reference_field_gaussian} against the
analytical result and comparison to some results that can be calculated by
other codes.

%
% case 1 Comparison ESRF spectrum_ebs_current%
% 
% \subsection{ESRF and EBS}
We performed a decomposition of the radiation from the 2~m long u18 ESRF undulator 
for its first harmonic ($E=7982$~eV, deflection parameter $K=1.68$) at its virtual 
source position placed in the current ESRF lattice
and in the future ESRF-EBS lattice (parameters from ~\cite{orangebook} summarized in the supplementary
material).

%
% Figure 1 mode_distribution/ebs_and_current
%

\begin{figure}
\begin{overpic}[width=8.5cm]
{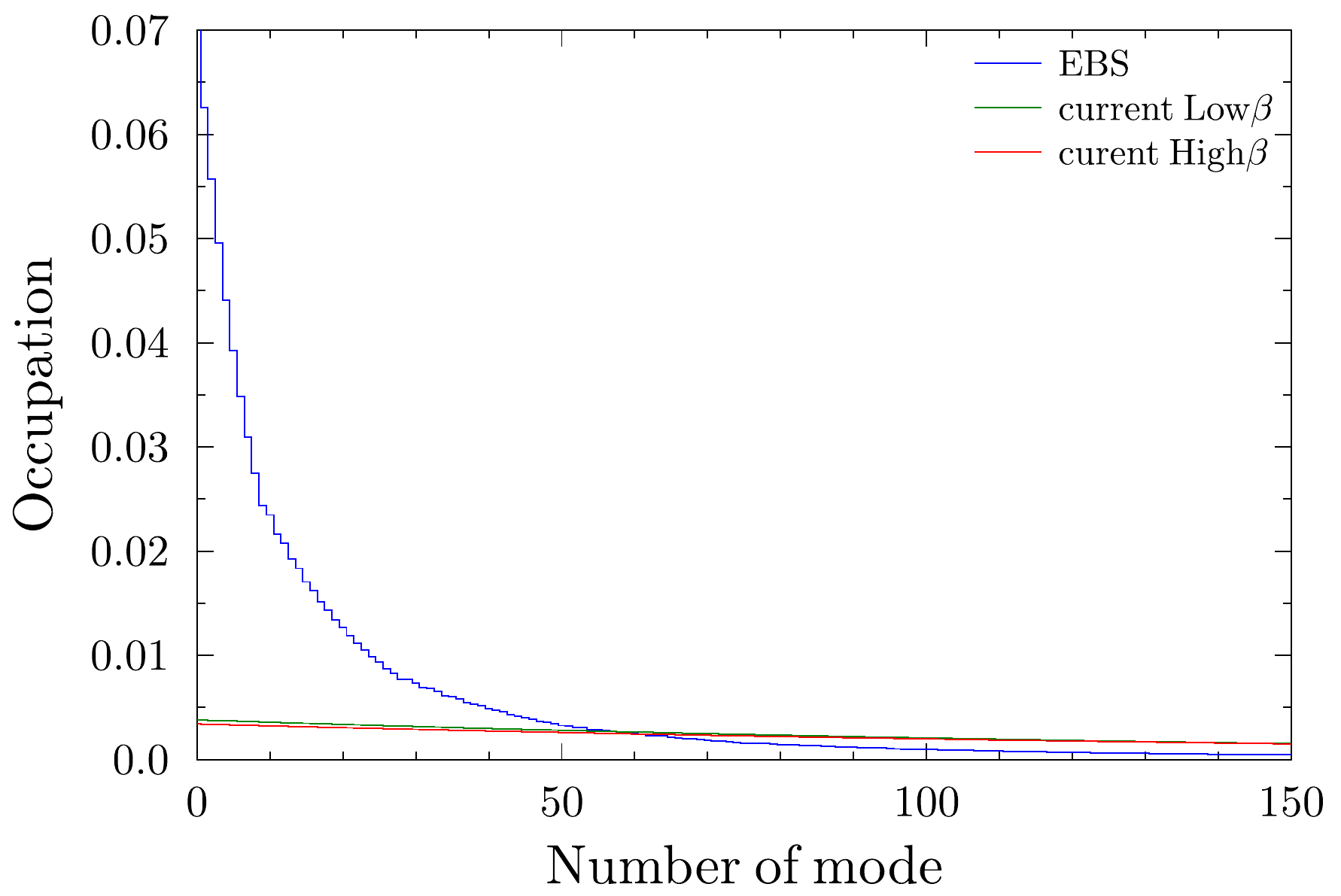}
\put(80,40){\includegraphics[width=4.5cm]{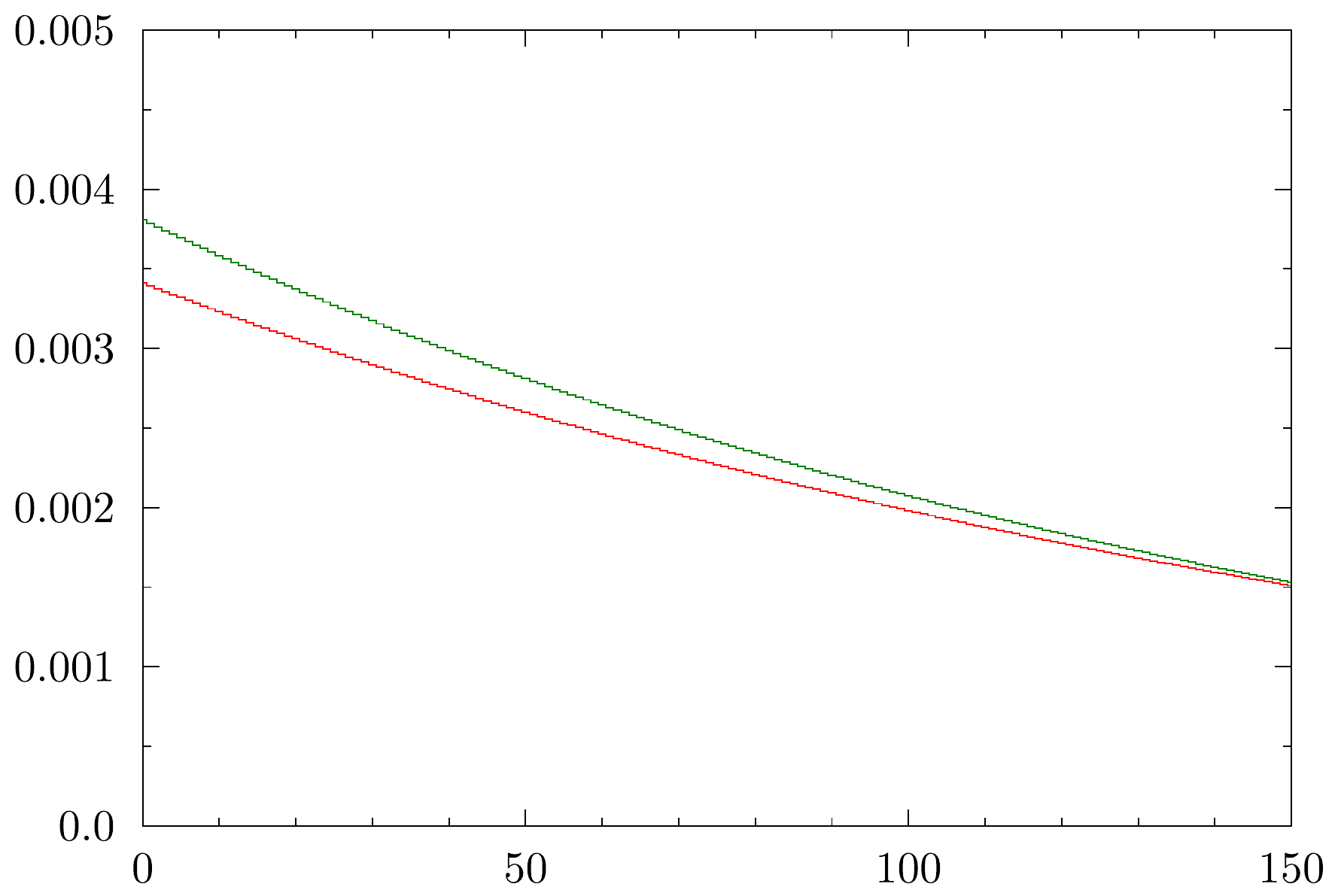}}
\end{overpic}
\caption{Eigenvalue spectra for the 2~m long ESRF u18 undulator at its first harmonic for the new ESRF-EBS lattice
and for the current ESRF lattice (high and low beta sections).
A detailed view of the spectra for the current ESRF lattice is the inset.}
\label{spectrum_ebs_current}
\end{figure}

The eigenvalue spectrum (\refpic{spectrum_ebs_current}) or mode
distribution gives direct information on how the beam spectral density
splits into different modes.
The $m$-th mode eigenvalue has an occupation $d_m$ defined by \refeq{spectrum}.
The occupation of the zeroth mode can be used to define a coherent
fraction ($CF=d_0$). In the limit $d_0\gg d_1$ most of the
spectral density is due to the first coherent mode and the contributions of
higher coherent modes to the spectral density can be considered as statistical
noise. We find that in the case of the ESRF-EBS the coherent fraction is an
order of magnitude higher than for the current ESRF lattice
(\refpic{spectrum_ebs_current}).

\begin{figure}
\begin{overpic}[width=8.5cm]
{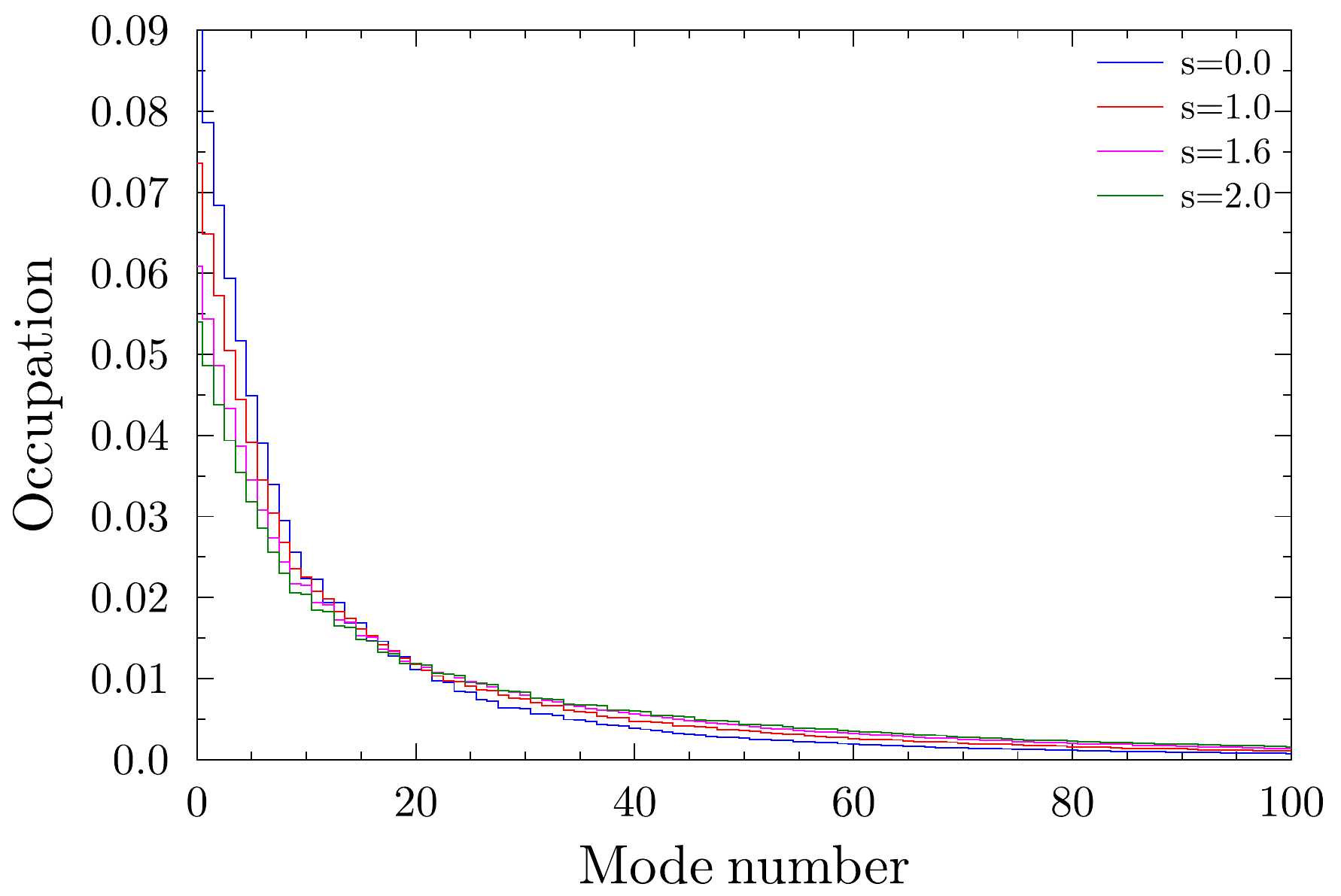}
\put(80,40){\includegraphics[width=4.5cm]{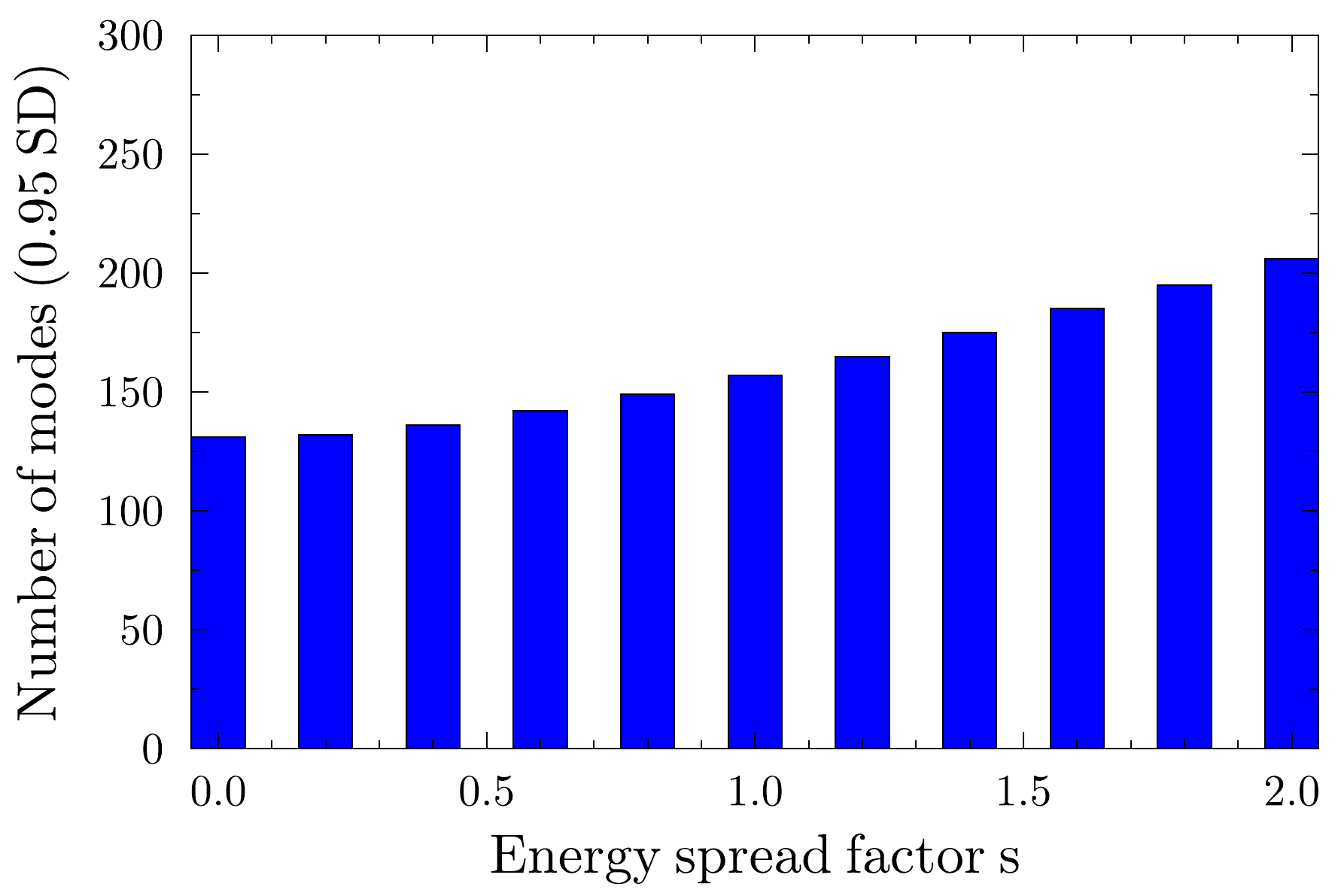}}
\end{overpic}
\caption{Number of modes for a few $s$ values of a 
$95\%$ of the spectral density and the mode distribution (inset) for a
linear variation of the energy spread $\left(\sigma_{\delta}\right)' = \sigma_{\delta}  \cdot s$ with a 4m long ESRF u18 undulator and
ESRF-EBS lattice settings($s=1.0$ corresponds to energy spread of 0.001).}
\label{compares_ces}
\end{figure}

Let us now discuss the effect of the electron beam energy spread on the mode
spectrum. We use ESRF-EBS lattice settings and an ESRF u18 undulator of 4~m
length at its first harmonic. The ESRF-EBS energy spread is varied according
to:
\[
\left(\sigma_{\delta}\right)' = \sigma_{\delta}  \cdot s,
\]
with $s\in\{0.0, 0.2,0.4,\ldots,1.0,\ldots, 1.8,2.0\}$,
whereas the transverse beam parameters are kept fixed. The energy integration was performed with $27$ points. 
We observe that the total number of coherent modes to incorporate $95\%$
of the spectral density are relatively largely changed (see
\refpic{compares_ces}). The minimal number of coherent modes is
$131$ which is achieved in the limit of vanishing energy spread. Doubling the
ESRF-EBS energy spread results in $206$ coherent modes. This is about $30\%$ more than
the $157$ modes for the present ESRF-EBS lattice settings.

%
% Figure 2 compare_energy_spread
%

% \twopictures{modes_95percent.eps}{ces_4m_distribution.eps}{7.5}{Number of modes for a few $s$ values of a 
% $95\%$ of the spectral density(left) and the mode distribution (right) for a
% linear variation of the energy spread $\left(\sigma_{\delta}\right)' = \sigma_{\delta}  \cdot s$ with a 4m long ESRF u18 undulator and
% ESRF-EBS lattice settings($s=1.0$ corresponds to energy spread of 0.001).}{compares_ces}

Our implementation of the algorithm can take energy spread into account and
works both at positions with zero Twiss alpha (symmetry point), and at positions with 
finite Twiss alpha as long as the vertical, horizontal and energy dimensions of the electron
beam are uncoupled.

% \subsection{A 1:1 imaging beamline}

\begin{figure}
\includegraphics[width=8.5cm]{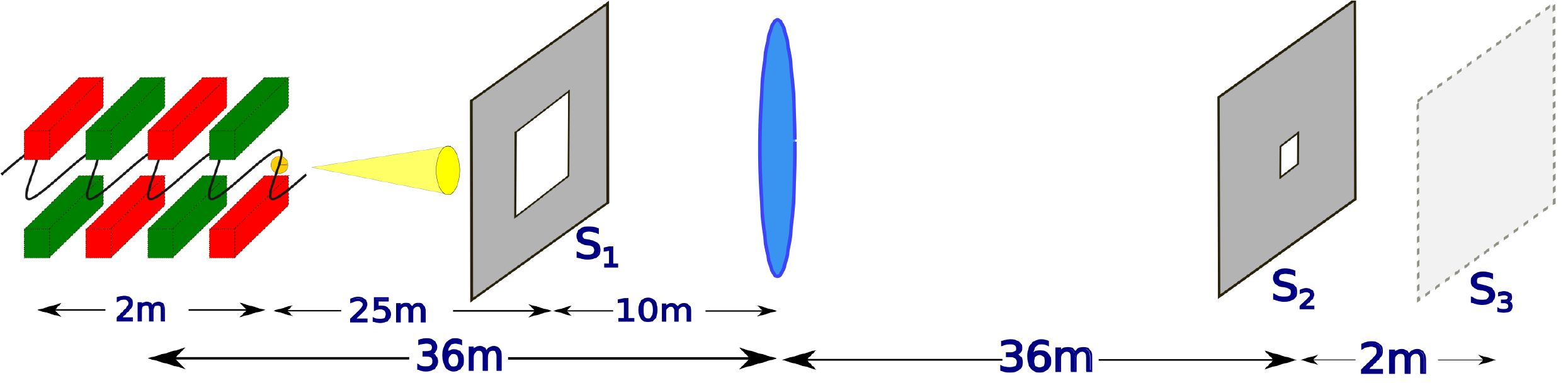}
\caption{Schematic view of the 1:1 imaging beamline of the undulator center.}
\label{one_to_one_beamline}
\end{figure}

We calculated an ideal 1:1 imaging beamline (\refpic{one_to_one_beamline}) at ESRF-EBS lattice settings.
A slit $S1$ collects the radiation from the undulator and is opened to accept the full central cone. 
A focusing system is idealized by an ideal lens focusing the virtual source (at 36 m upstream the
lens) into the $S2$ plane (at 36 m downstream the lens). The decomposition of the
radiation in coherent modes at $S1$ presents the same spectrum shown before \refpic{spectrum_ebs_current}. 
The spectrum calculated at $S2$ is the same as the spectrum in $S1$, 
because the lens is an ideal element that modifies the individual eigenfunctions but conserves the 
integral of the spectral density.
The CSD can be calculated from the coherent modes. It is a function of four
coordinates. We plot horizontal and vertical cuts in \refpic{two_pictures_compares_ces1}. The full-width at half maximum of the profile
with one coordinate fixed at zero gives the coherence length of $52.9~\mu$m in
the horizontal and of in $62.4~\mu$m the vertical.
In the supplementary material these CSD results for a different beamline are
compared with SRW Monte-Carlo calculations showing a very good agreement.
The first two coherent modes are shown in  \refpic{two_pictures_beamline_modes}.

%
% Figure 3 beamline
%

% \onepicture{bl_one_to_one.pdf}{14}
% {Schematic view of the 1:1 imaging beamline of the undulator
% center.}
% {one_to_one_beamline}

%
% Figure 4 CSD
%

\twopictures{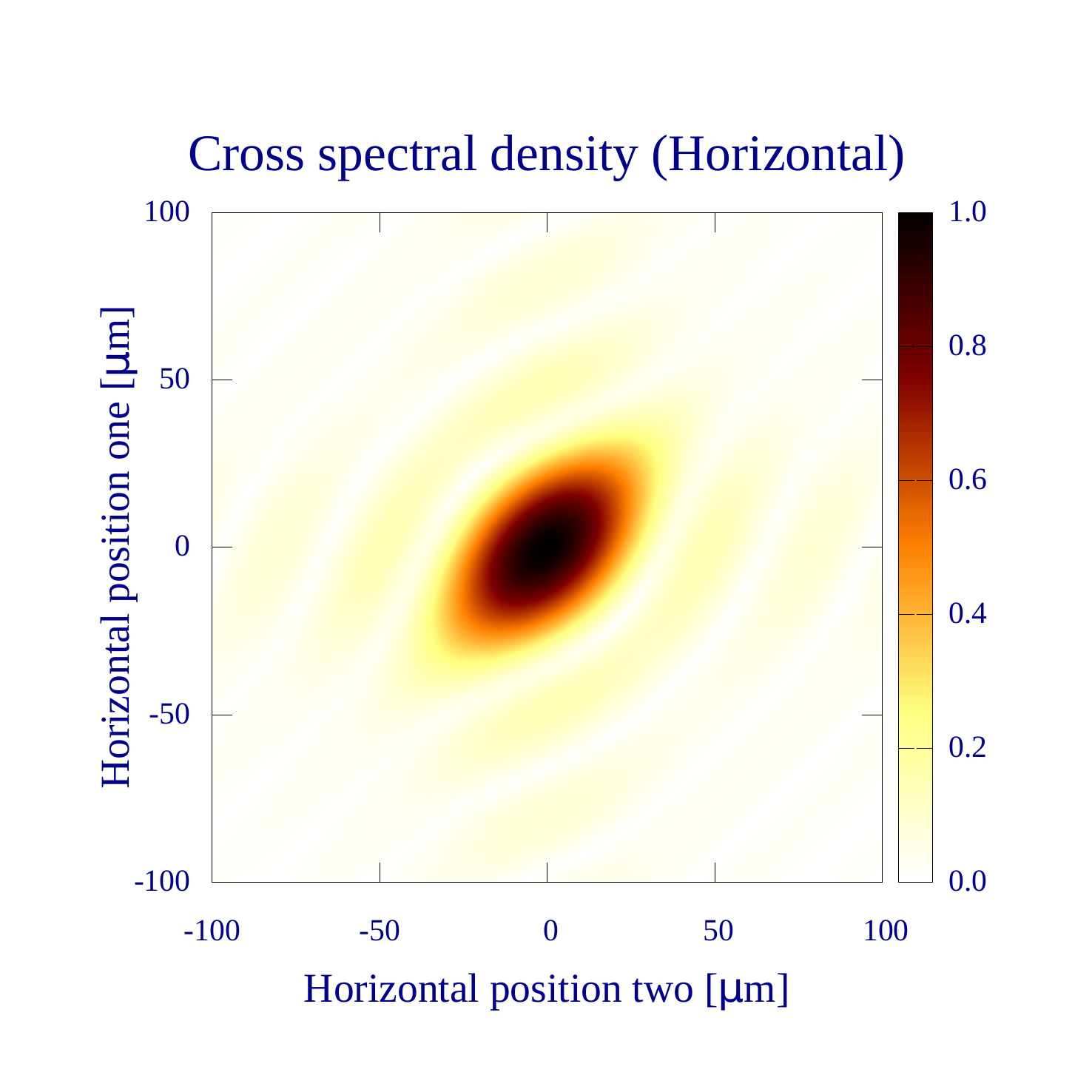}{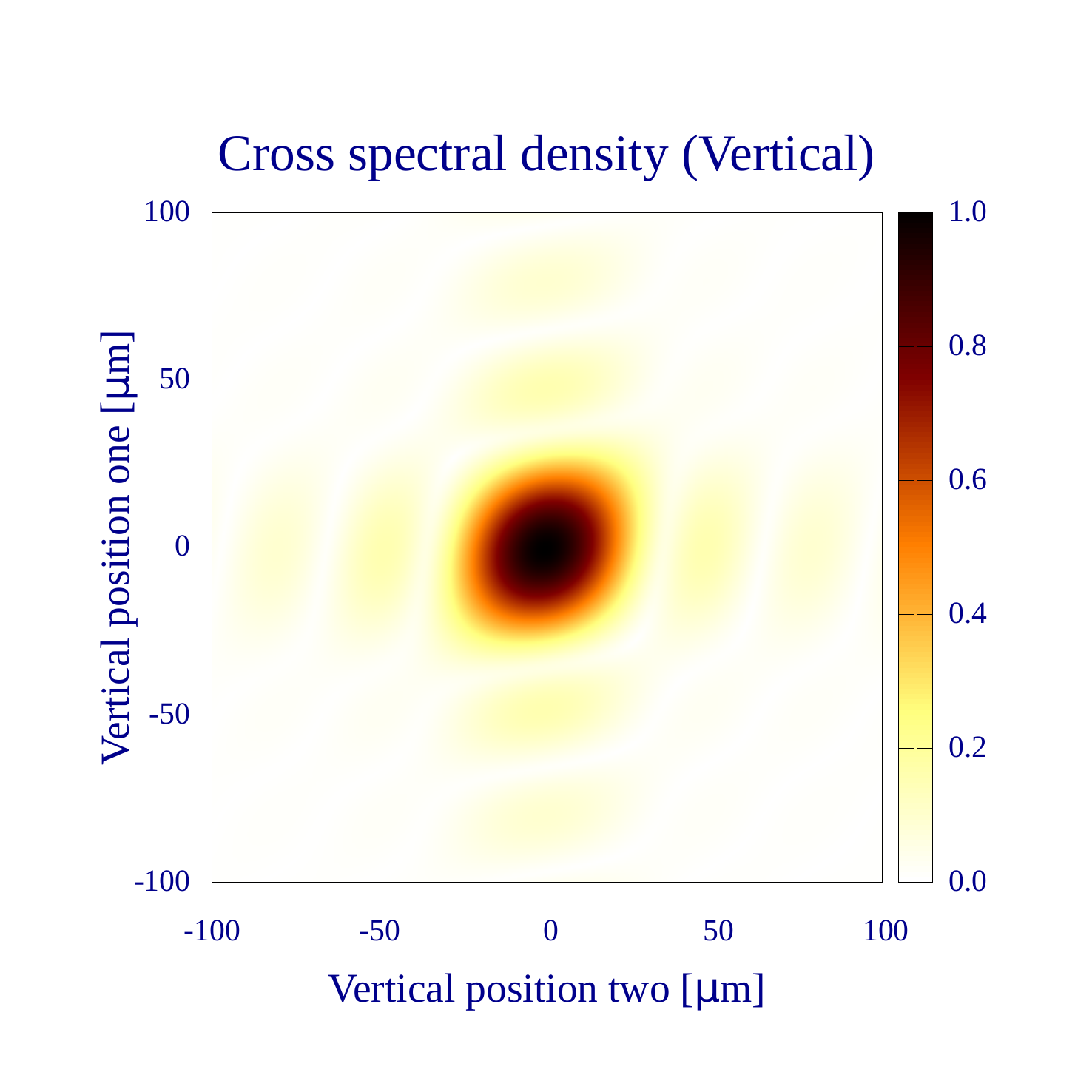}{5}
{Vertical and horizontal cuts, i.e. the coordinates of the other dimension are
fixed at zero, of the normalized cross spectral density with slit size
$5\mu m\times 5 \mu m$ at $S_2$ calculated from coherent modes for a 2m long
ESRF u18 undulator at the ESRF-EBS lattice at $S_3$ for the beamline shown in
\refpic{one_to_one_beamline}.}{compares_ces1}

%
% Figure 5 Modes
%

\twopictures{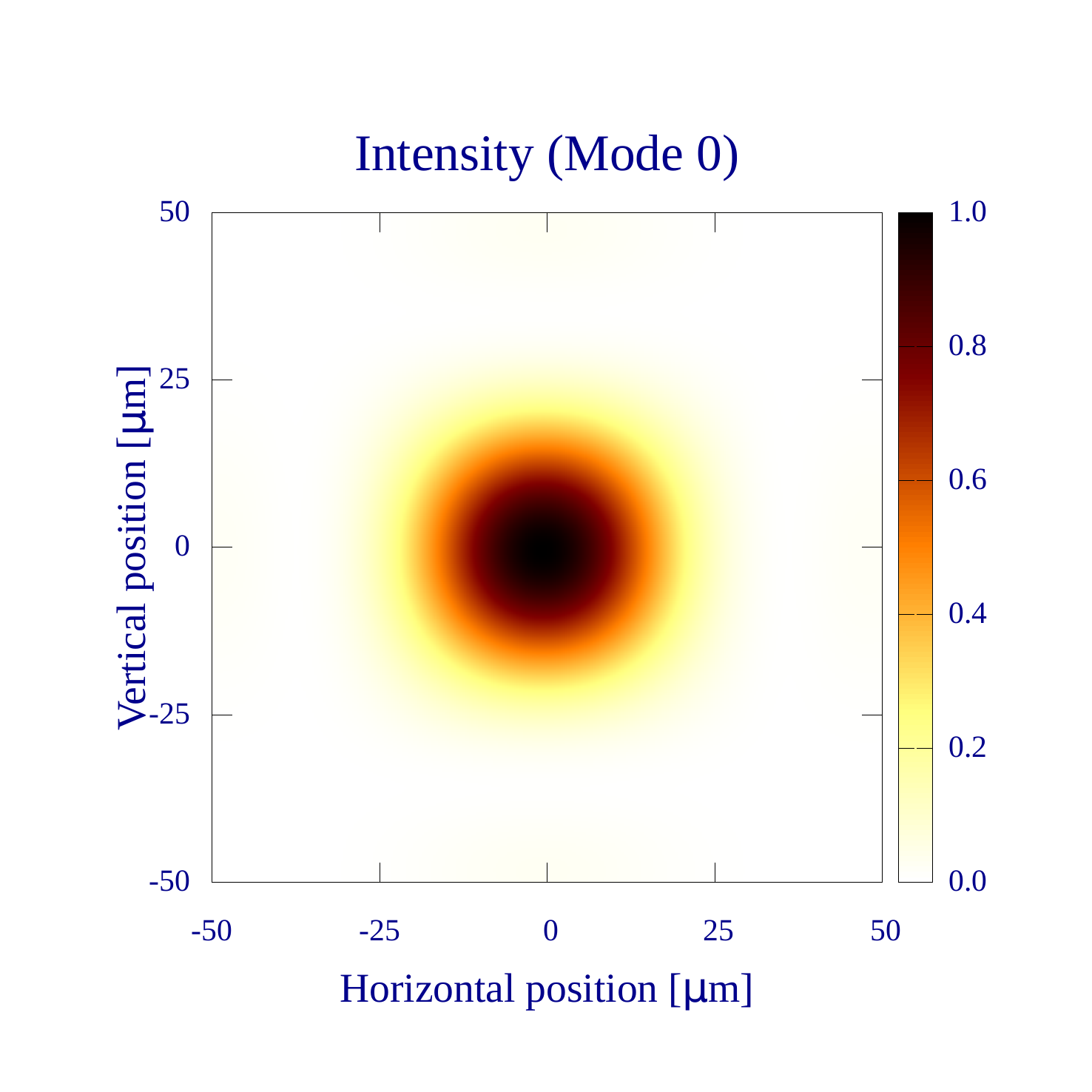}{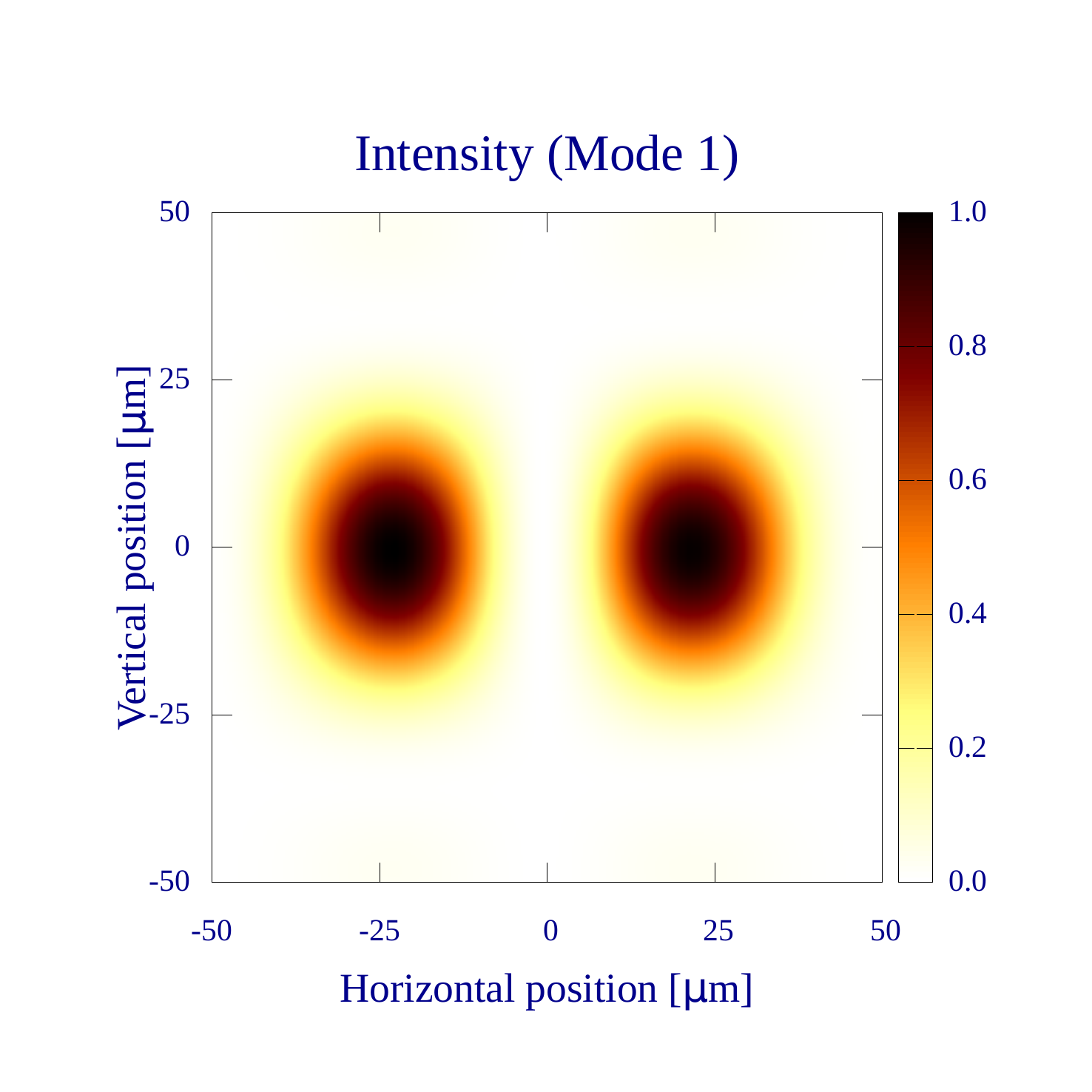}{5}
{Zeroth and first coherent modes at $S_3$ for the beamline shown in
\refpic{one_to_one_beamline} with a $5\mu m\times 5\mu m$ aperture
size at $S_2$} {beamline_modes}

To mimic in an idealized fashion the experiment made in \cite{Pelz2014} we have
at the $S2$ plane an aperture with varying
size.
After the slit aperture we propagate by an additional 2~m in free space to the new final
position $S3$. At the final position we perform another coherent mode decomposition. We find that 
reducing the aperture size leads to a narrowing of the eigenvalue spectrum with a
simultaneous decrease of spectral density at the final position \refpic{two_pictures_slit_variation}.
We observe that closing the slit improves the overall coherence for the price of spectral density.
In the limit $d_0\gg d_1$ almost the entire spectral density is
carried by the first coherent mode. The other modes may be regarded as statistical noise.

%
% Figure 6 aperture
%

\twopictures{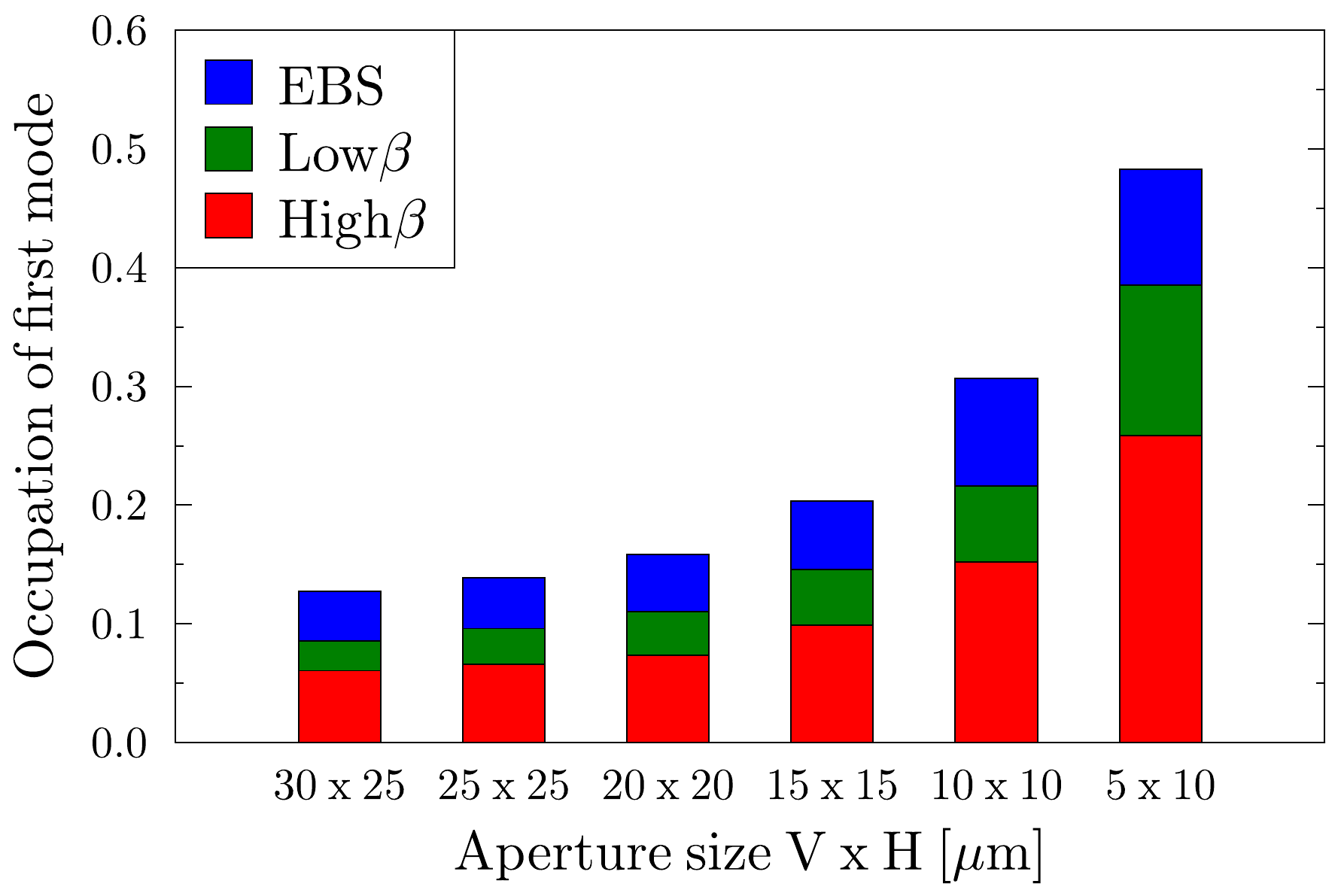}{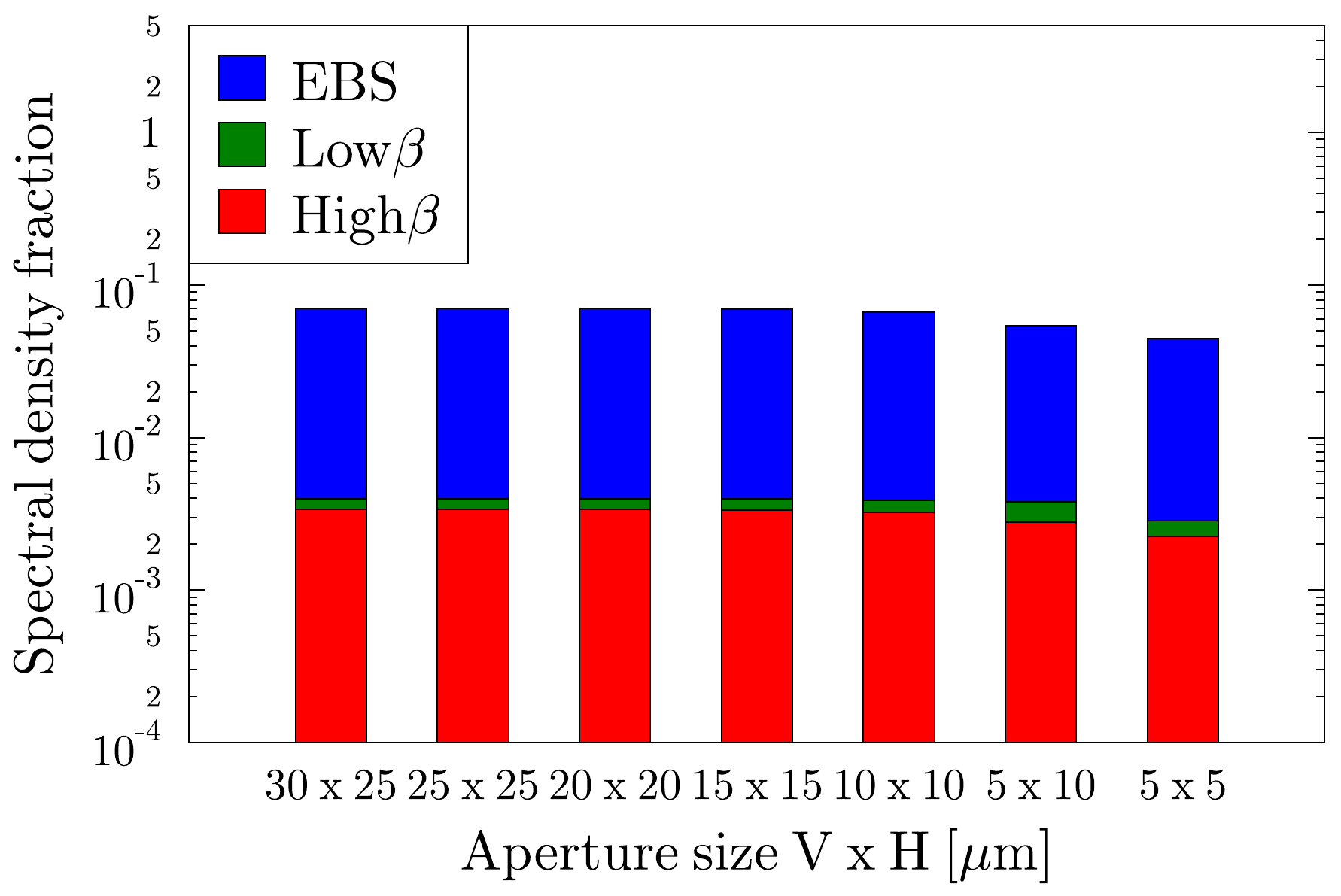}{4.5}{First
mode occupation (left) and spectral density fraction(right) of the first mode at
$S_3$ for a given slit size (VxH in $\mu m$) at $S_2$ for the beamline shown in
\refpic{one_to_one_beamline}. The fraction is calculated
with respect to the spectral density at the undulator virtual source. }{slit_variation}

% \section{Summary}
In summary we present an algorithm for the numerical coherent mode
decomposition for storage ring undulator cross spectral densities. Our
implementation of this algorithm can calculate the cross spectral density for
positions where the horizontal, vertical and electron energy dimensions of the
electron beam are uncoupled. The effect of energy spread and
finite Twiss alpha can be taken into account. The eigenvalue spectrum allows to define a quality
measure of coherence if the eigenvalue spectrum is dominated by the first
coherent mode. We propose a coherence fraction $\cohf$ as the fraction of the
largest eigenvalue over the integrated spectral density.
For low emittance storage ring emission, the number of coherent modes is small and we may propagate 
the cross spectral density to any point in the beamline efficiently by the
propagation of the coherent modes.
Any simulation that can model the  propagation of a single
electron emission along the beamline can also be used for the propagation
of the coherent modes and then to build the cross spectral density.
Additionally to the applications discussed in this paper, the coherent modes
could serve as an initial guess for reconstruction algorithms in coherent diffraction imaging techniques. 
Furthermore, it can be used for modeling X-ray-sample interactions: the coherent modes can be
propagated up to the sample position; the action of the sample is applied
directly to the propagated modes. This is more efficient than a
multi-electron Monte Carlo approach because the optical elements before the sample do not need to be 
recalculated and the number of coherent modes is in general much smaller than the 
number of simulated electrons in the Monte Carlo simulation.
The same applies to the design of beamlines, where after every design change
only a small number of coherent modes after the last modified optical element have to be propagated.
The code of our algorithm is open-source and can be found at \cite{code}.

% \bibliography{references}
% \bibliographystyle{ieeetr}

% \end{document}

\end{document}